\documentclass{aa}
\usepackage{graphicx}
\usepackage{txfonts}

\begin{document}

\title{XMM-Newton observation of the most X-ray-luminous galaxy cluster 
RX J1347.5$-$1145}

\author{Myriam Gitti
 \and Sabine Schindler}

\institute{Institut f\"ur Astrophysik, Leopold-Franzens Universit\"at 
Innsbruck, Technikerstra\ss e 25, A-6020 Innsbruck, Austria}

\offprints{Myriam Gitti,
\email{myriam.gitti@uibk.ac.at}}

\authorrunning{Gitti \& Schindler.}
\titlerunning{XMM-Newton observation of the most X-ray-luminous
galaxy cluster RX J1347.5$-$1145 }
\date{Received / Accepted}

\abstract{
We report on an XMM-Newton observation of  RX J1347.5$-$1145 
(z=0.451), the most luminous X-ray cluster of galaxies 
currently known, with a luminosity $L_X = 6.0 \pm 0.1 \times 10^{45}$ erg/s  
in the [2-10] keV energy band. 
We present the first temperature map of this cluster, which shows a
complex structure. It identifies the cool core and a hot region at 
radii 50-200 kpc to south-east of the main X-ray peak, at a position 
consistent with the subclump seen in the X-ray image. This structure is
probably an indication of a submerger event.
Excluding the data of the south-east quadrant, the cluster appears relatively
relaxed and we estimate a total mass within 1.7 Mpc of 2.0$\pm 0.4 
\times 10^{15} M_{\sun}$. We find that the overall temperature of the cluster 
is $kT=10.0 \pm 0.3$ keV. 
The temperature profile shows a decline in the outer regions
and a drop in the centre, indicating the presence of a cooling core which can
be modelled by a cooling flow model with a minimum temperature $\sim$2 keV
and a very high mass accretion rate,
$\dot{M} \sim 1900 \mbox{ M}_{\sun}/ \mbox{yr}$. 
We compare our results with previous observations from \textit{ROSAT}, 
\textit{ASCA} and \textit{Chandra}.
}
\maketitle


\section{Introduction}

In this paper we present the first results from an \textit{XMM-Newton} 
observation of RX J1347.5$-$1145, 
the most X-ray-luminous galaxy cluster known (Schindler et al. 1995).
This cluster has been detected in the \textit{ROSAT} All-Sky Survey and further
studied with \textit{ROSAT} HRI and \textit{ASCA} (Schindler et al. 1995, 
1997).
It shows a very peaked X-ray emission profile and presents a strong cooling 
flow in its central region. Submm observations in its direction 
showed a very deep SZ decrement (Komatsu et al. 1999, 2001; 
Pointecouteau et al. 1999, 2001).
Due to the presence of gravitational arcs, this cluster is also well suited 
for a comparison of lensing mass and X-ray mass. 
Optical studies of weak lensing have been performed by Fischer \& Tyson (1997)
and Sahu et al. (1998).   
Recent \textit{Chandra} observations (Allen et al. 2002) discovered  a 
region of relatively hot, bright X-ray emission, located approximately 20 
arsec to the south-east of the main X-ray peak at a position consistent
with the region of enhanced SZ effect. This could be an indication for 
a subcluster merger, pointing to a complex dynamical evolution.
A comparison of the \textit{XMM-Newton} and SZ results, a more detailed 
analysis of the complex dynamical state of the cluster and a comparison
of lensing mass and X-ray mass will be presented in a forthcoming
paper (Gitti et al. in prep.).

RX J1347.5$-$1145 is at a redshift of 0.451. With 
$H_0 = 70 \mbox{ km s}^{-1} \mbox{ Mpc}^{-1}$, and 
$\Omega_M=1-\Omega_{\Lambda}=0.3$, 
the luminosity distance is 2506 Mpc and 
1 arcsec corresponds to 5.77 kpc.


\section{Observation and data preparation}

RX J1347.5$-$1145
 was observed by \textit{XMM-Newton} in July 2002 during rev. 
484 with the MOS and pn detectors in Full Frame Mode with THIN filter.
We used the SASv6.0.0 processing tasks \textit{emchain} and \textit{epchain} to
generate calibrated event files from raw data.
Standard processing was applied to prepare data and reject the soft proton
flares. The remaining exposure times after cleaning are 32.2 ks for MOS1,
32.5 ks for MOS2 and 27.9 ks for pn. 
The background estimates were obtained using a blank-sky observation consisting
of several high-latitude pointings with sources removed (Lumb et al. 2002).
The background subtraction 
was performed as described in full detail in Arnaud et al. (2002).
The source and background events were corrected for vignetting using the
weighted method described in Arnaud et al. (2001).


\section{Morphological analysis}
\label{morphology.sec}

\begin{figure}[ht]
\includegraphics{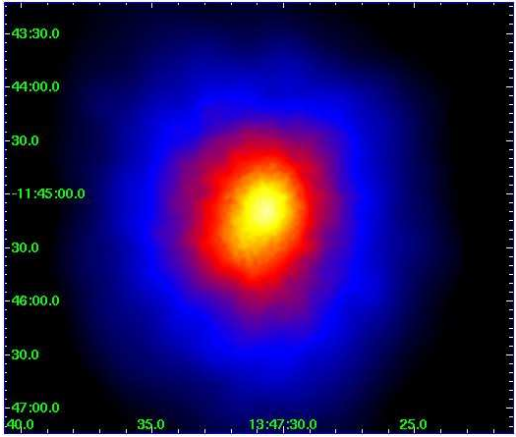}
\vspace{5.5cm}
\caption{
Total (MOS+pn) \textit{XMM-Newton} EPIC mosaic image of 
RX J1347.5$-$1145
in the [0.9-10] keV energy band. The image is corrected for vignetting and
exposure and is adaptively smoothed.}
\label{mosaic}
\end{figure}

\begin{figure}[ht]
\includegraphics{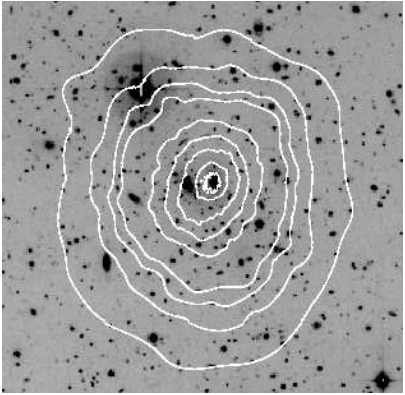}
\vspace{5cm}
\caption{
VLT image of the central region of RX J1347.5$-$1145 (Erben et al. in 
prep.). Superposed are the ([0.9-10] keV) XMM X-ray contours (levels: 
0.003, 0.015, 0.045, 0.06, 0.15, 0.3, 0.6, 1.5, 3 cts/s/arcmin$^{2}$).
The image is $\sim$ 4.4 $\times$ 4.3 arcmin$^2$ (North is up,
East is left).
}
\label{optical}
\end{figure}

The adaptively smoothed, exposure corrected image (MOS+pn) 
in the [0.9-10] keV energy band is presented in Fig. \ref{mosaic}. 
The image is obtained from the mosaic of the raw images corrected for the 
mosaic of the exposure maps by running the task
\textit{asmooth} set to a desired signal-to-noise ratio of 20. 
A number of notable features are visible. 
In particular, we note a sharp central surface brightness peak at 
$13^{\rm h} 47^{\rm m} 30^{\rm s}.6 -11^{\circ} 45' 09''.0$ (J2000), in very 
good agreement with the optical centroid for the dominant cluster galaxy 
(Schindler et al. 1995). We also confirm the presence of a region of
enhanced emission $\sim$ 20 arcsec to the south-east (SE) of the X-ray peak 
and, on large scale ($\sim$ 80 arcsec), the extension of the X-ray emission to 
the south, already revealed in previous observations with
\textit{Chandra} (Allen et al. 2002).
In Fig. \ref{optical} we show an overlay of the VLT image of the central 
region of RX J1347.5$-$1145 with the X-ray contours derived from 
Fig. \ref{mosaic}.

\begin{figure}
\includegraphics{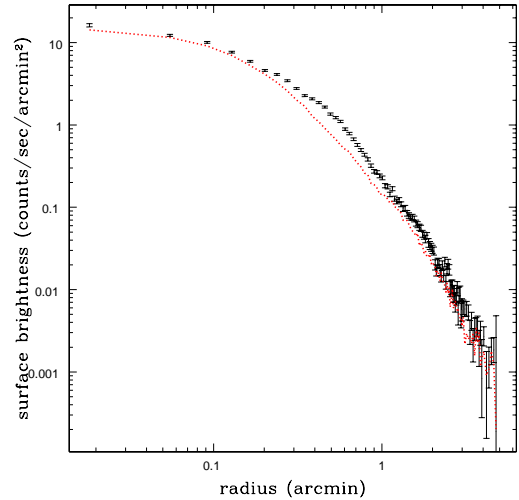}
\vspace{6.5cm}
\caption{
Background subtracted, azimuthally averaged radial surface brightness profile 
for SE quadrant data in the [0.3-2] keV range. The dotted line shows the 
profile in other directions (undisturbed cluster), which appears relatively 
regular and relaxed. An excess of emission in the SE quadrant
between radii of $\sim$ 20-50 arcsec (100-300 kpc) is visible.}
\label{profile-SE}
\end{figure}

We compute a background-subtracted vignetting-corrected radial surface
brightness profile in the [0.3-2] keV energy band for each camera separately.
The profiles for the three detectors are then added into a single profile,
binned such that at least a sigma-to-noise ratio of 3 was reached.
The cluster emission is detected up to 1.7 Mpc ($\sim 5'$). 
In Fig. \ref{profile-SE} we show the X-ray surface brightness profiles for 
the disturbed SE quadrant compared to that from
data excluding the SE quadrant. We note that the data excluding the 
SE quadrant (hereafter undisturbed cluster) appear regular, 
while those for the SE quadrant (containing the X-ray 
subclump) show a clear excess of emission between radii of $\sim$ 100 and 300
kpc relative to other directions.  

The surface brightness profile of the undisturbed cluster is fitted in the 
CIAO tool \textit{Sherpa} with various 
models, which are convolved 
with the \textit{XMM-Newton} PSF. 
A single $\beta$-model (Cavaliere \& Fusco Femiano 1976) is not a good
description of the entire profile: a fit to the outer regions (350 kpc - 
1.7 Mpc) shows a strong excess in the centre when compared to the
model. 
The peaked emission is a strong indication for a cooling core in this cluster.
We found that for 350 kpc-1.7 Mpc the data can be described by a 
$\beta$-model with a core radius $r_{\rm c}=367 \pm 3$ kpc and a slope
parameter $\beta=0.93 \pm 0.01$, while for $r<$350 kpc
the data can be approximated by a $\beta$-model with $r_{\rm c}=40 \pm 0.2$ 
kpc and $\beta=0.55 \pm 0.02$ (90\% confidence levels).


\section{Temperature map}

\begin{figure}[ht]
\includegraphics{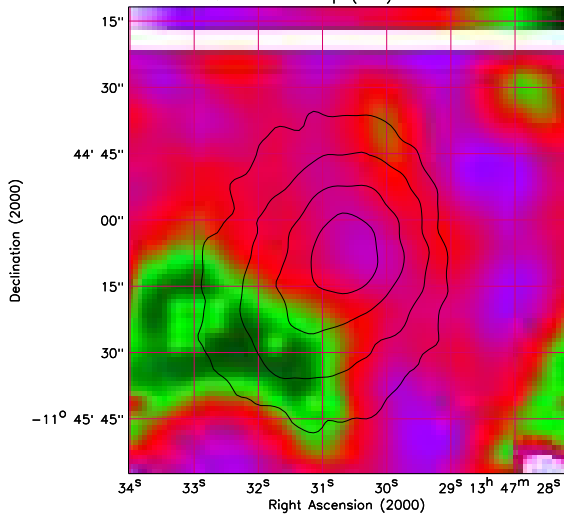}
\vspace{6.5cm}
\caption{
Temperature map obtained by using 4 X-ray colours ([0.3-1], [1-2], [2-4.5],
[4.5-8] keV) and estimating the expected count rate with XSPEC for a thermal
MEKAL model, with fixed Galactic absorption $N_{\rm H} = 4.85 \times 10^{20} 
\mbox{ cm}^{-2}$ and metallicity $Z = 0.3 Z_{\sun}$. 
Superposed are the X-ray contours. The features outside the last contours are 
not significant, as they are mainly due to noise fluctuations.
}
\label{tmap.fig}
\end{figure}

The temperature image of the central cluster region shown in Fig. 
\ref{tmap.fig} is build from X-ray colours. Specifically, we produce the 
mosaics of MOS images in four different energy bands 
([0.3-1] keV, [1-2] keV, [2-4.5] keV and [4.5-8] keV), 
subtract the background and divide the resulting
images by the exposure maps. A temperature is 
obtained by fitting the values in each pixel with a thermal 
plasma. In particular we note that the very central region
appears cooler than the surrounding medium and the SE quadrant,
which corresponds to the subclump seen in the X-ray image,
is significantly hotter than the gas in other directions. 

\section{Spectral analysis}
\label{spectral.sec}

For the spectral analysis we treat the SE quadrant containing 
the X-ray subclump separately from the rest of the cluster. 
The data for the undisturbed cluster are divided into the annular regions 
detailed in Table 1. A single spectrum is extracted for each region and 
then regrouped to contain a minimum of 25 counts per channel, 
thereby allowing $\chi^2$ statistics to be used. 
The data from the three cameras were modelled simultaneously using the XSPEC
code, version 11.3.0. Spectral fitting is performed in the [0.5-8] keV band.
The spectra are modelled using a simple, single-temperature model
(MEKAL plasma emission code in XSPEC) with the absorbing column density 
fixed to the Galactic value ($N_{\rm H} = 4.85 \times 10^{20} 
\mbox{ cm}^{-2}$, Dickey \& Lockman 1990). The free parameters in this
model are the temperature $kT$, metallicity $Z$ (measured relative to the
solar values) and normalization (emission measure).

\begin{figure}[ht]
\includegraphics{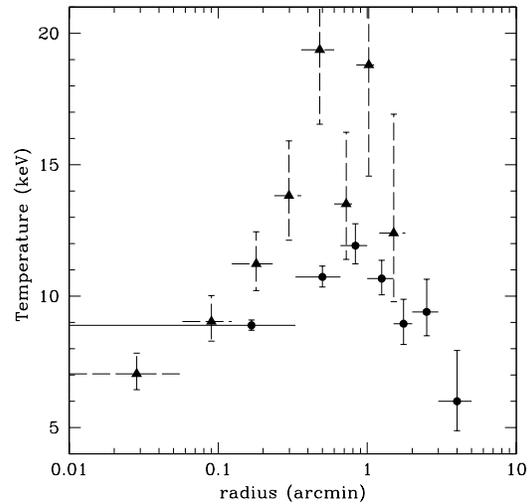}
\vspace{6.5cm}
\caption{
\textit{Circles}: the projected X-ray gas temperature profile (and errors at 
90\% confidence levels) measured from \textit{XMM} data in the [0.5-8] keV 
energy band. 
\textit{Triangles}: the projected X-ray gas temperature profile (and 1$\sigma$
errors) measured from \textit{Chandra} data in the [0.5-7] keV energy 
band (Allen et al. 2002). The data of the SE quadrant are 
excluded in both profiles.}
\label{profilot.fig}
\end{figure}

The best-fitting parameter values and 90\% confidence levels derived from the
fits to the annular spectra are summarized in Table 1.
The projected temperature profile determined with this model is shown
in Fig. \ref{profilot.fig}. The temperature rises from a mean value of
$8.8 \pm 0.3$ keV within 115 kpc to $kT = 11.1 \pm 0.4$ keV over the 0.1-0.5
Mpc region, then declines down to a mean value of $6.0^{+2.6}_{-1.6}$ keV 
in the outer regions (1.0-1.7 Mpc). 
In Fig. \ref{profilot.fig} we also show for comparison the projected 
temperature profile measured by \textit{Chandra} (Allen et al. 2002). 
We note that while the
general trend observed by the two satellites is consistent, there are some
discrepancies in the measurements of the absolute temperature values.  
The discrepancy between \textit{Chandra} and \textit{XMM} temperature profile
has been found in other clusters of galaxies (e.g. A1835, Schmidt et al. 2001,
Majerowicz et al. 2002), and can be partially due to the effect of the 
\textit{XMM} PSF (see Markevitch 2002).
\begin{table}
\small
\caption
{\small
The results from the spectral fitting in concentric annular regions 
(undisturbed cluster). Temperatures ($kT$) are in keV, metallicities ($Z$)
in solar units and [2-10] keV luminosities ($L_{\rm X}$) in units of 
$10^{44} \mbox{ erg s}^{-1}$.
 The total $\chi^2$ values and numbers of degrees of freedom 
(DOF) in the fits are listed in column 5. Errors are 90\% 
confidence levels ($\Delta \chi^2 = 2.71$) on a single parameter of interest. 
}
\hskip 0.2truein
\begin{tabular}{ccccc}
\hline  
Radius (kpc)   &     $kT$    & $Z$    & $L_{\rm X}$  &   $\chi^2$/DOF     \\
\hline
0-115 & $8.9^{+0.3}_{-0.3}$ & $0.34^{+0.05}_{-0.05}$ & 16.7 & 982/880 \\
115-230 & $10.7^{+0.7}_{-0.6}$ & $0.26^{+0.08}_{-0.08}$ &10.1 & 696/664 \\
230-345 & $11.9^{+1.6}_{-1.3}$ & $0.16^{+0.14}_{-0.15}$ & 5.44 &  433/384 \\
345-520 & $10.7^{+1.1}_{-1.0}$ & $0.24^{+0.13}_{-0.13}$ & 4.52&  350/341 \\
520-690 & $9.0^{+1.4}_{-1.1}$ & $0.16^{+0.18}_{-0.16}$ & 1.92& 239/210 \\
690-1040 & $9.4^{+2.1}_{-1.4}$ & $0.19^{+0.26}_{-0.19}$ & 1.64& 293/264 \\
1040-1730 & $6.0^{+2.6}_{-1.7}$ & $0.40^{+0.50}_{-0.37}$ & 0.91& 593/421 \\
0-1730 & $9.4^{+0.3}_{-0.3}$ & $0.26^{+0.04}_{-0.04}$ & 41.8& 1957/1452 \\
\hline                                   
\end{tabular}
\label{profilot.tab}
\end{table}
A fit with the same model to the data for the SE quadrant between
radii 50-200 kpc yields a best-fitting temperature $kT=13.3 \pm 1.0$ keV. 
In other directions, the mean value is $kT=11.0^{+0.5}_{-0.4}$ keV. 
The metallicity profile derived with the single-temperature model is 
consistent with being constant, with an overall value of $Z=0.26 \pm 0.04 
Z_{\sun}$. 
However, as shown in Table 2, the structure in the innermost bin 
is better modelled by multi-temperature models having higher metallicities.

Within the radius of 1.7 Mpc ($\sim 5'$), a fit to the full 360$^{\circ}$ data
gives an overall $kT = 10.0 \pm 0.3$ keV, $Z=0.26 \pm 0.03 Z_{\sun}$ and 
$L_X \mbox{ (2-10 keV)}= 6.0 \pm 0.1 \times 10^{45} \mbox{ erg s}^{-1}$. 
These values are in agreement with \textit{ROSAT} and
\textit{ASCA} results (Schindler et al. 1997).


\section{Mass determination}

\begin{figure}[ht]
\includegraphics{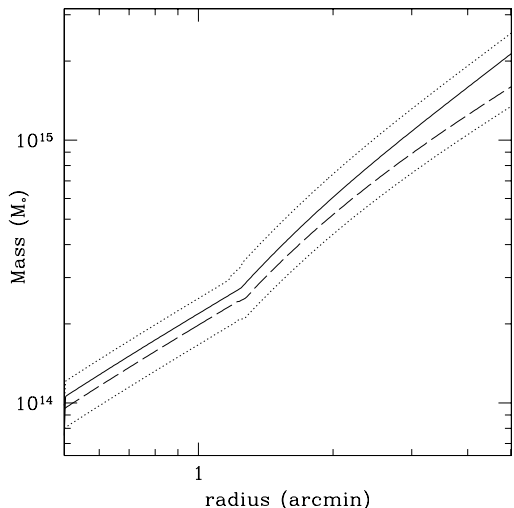}
\vspace{6.5cm}
\caption{
\textit{Solid line}: Profile of the integrated total mass.
\textit{Dashed line}: Profile of the integrated total mass calculated assuming
a constant temperature of 9.5 keV.
\textit{Dotted line}: Error on the mass calculation coming from the temperature
measurement.
}
\label{massa.fig}
\end{figure}

The total gravitating mass distribution shown in Fig. \ref{massa.fig} 
(solid line) was calculated under the usual assumptions of hydrostatic 
equilibrium and spherical symmetry 
using the deprojected density distribution calculated from
the parameters of the $\beta$-model derived in Sect. \ref{morphology.sec}.
Only data beyond $30''$ ($\sim 175$ kpc) are considered: in the central
bins the temperature as estimated in Sect. \ref{spectral.sec} is 
affected by the \textit{XMM} PSF and projection effects, while for the
outer regions these effects can be neglected (e.g. Kaastra et al. 2004). 
Within 1 Mpc we find a total mass of $1.0 \pm 0.2 \times 10^{15} M_{\sun}$, 
in agreement with \textit{Chandra} (Allen et al. 2002) and weak lensing 
analysis (Fischer \& Tyson 1997) results and
slightly higher than that derived by \textit{ROSAT/ASCA} 
(Schindler et al. 1997). 
In Fig. \ref{massa.fig} we also show for comparison (dashed line) 
the mass profile derived by assuming a constant temperature of 9.5 keV.

\section{Cooling core analysis}

We accumulate the spectrum in the central 30$^{''}$ ($\sim$175 kpc) by
excluding the data for the SE quadrant.
We use three different spectral models.
Model A is the MEKAL model already used in Sect. \ref{spectral.sec}.
Model B includes a single temperature component plus an isobaric multi-phase
component (MEKAL + MKCFLOW in XSPEC), where the minimum temperature, 
$kT_{\rm low}$, and the normalization of the multi-phase component, 
Norm$_{\rm low} = \dot{M}$,
are additional free parameters.
Finally, in model C the constant pressure cooling flow is replaced by a 
second isothermal emission component (MEKAL + MEKAL in XSPEC). As for model
B, this model has 2 additional free parameters with respect to model A: 
the temperature, $kT_{\rm low}$, and the normalization, 
Norm$_{\rm low}$, of the second component.
\begin{table}
\small
\caption
{
The best-fit parameter values and 90\% confidence limits from the spectral 
analysis in the central 0-30$^{''}$ region. 
Temperatures ($kT$) are in keV, metallicities ($Z$) as a fraction of the solar 
value 
and normalizations in units of $10^{-14} n_{\rm e} n_{\rm p} V / 4 \pi
D_{\rm A} (1+z)^2$ as done in XSPEC (for the MKCFLOW model the normalization
is parameterized in terms of the mass deposition rate $\dot{M}$, in 
$\mbox{M}_{\sun} \mbox{ yr}^{-1} $).}
\hskip 0.2truein
\begin{tabular}{cccc}
\hline  
Par.   &     Mod. A    & Mod. B    &     Mod. C    \\
\hline
$kT$ & $9.2^{+0.3}_{-0.3}$ &  $23.8^{+6.1}_{-4.7}$   & $17.7^{+5.7}_{-3.6}$ \\
$Z$ & $0.32^{+0.05}_{-0.05}$ &  $0.39^{+0.06}_{-0.06}$  & $0.42^{+0.06}_{-0.06}$ \\  
Norm & 0.00528$^{+0.00009}_{-0.00008}$ & 0.00053$^{+0.00223}_{-0.00053}$  &  0.00355$^{+0.00051}_{-0.00048}$ \\ 
$kT_{\rm low}$ &  --- &  $2.0^{+0.5}_{-0.4}$   & $3.9^{+0.7}_{-0.6}$ \\ 
Norm$_{\rm low}$   & --- & ${\dot M}=1880^{+260}_{-210}$ &  0.00194$^{+0.00057}_{-0.00058}$ \\  
$\chi^2$/DOF & 1048/1003 &  1011/1001  & 1010/1001 \\ 
\hline                                   
\end{tabular}
\label{cf.tab}
\end{table}
The results, summarized in Table 2, show that the statistical 
improvements obtained by introducing an additional emission component 
(models B or C) compared to the single-temperature model (model A) are 
significant at more than the 99\% level according to the F-test,
although the temperature of the hot gas is unrealistically 
high.
With our data, however, we cannot distinguish between the two multi-phase 
models. This means that the extra emission component can be equally well 
modelled either as a cooling flow or a second isothermal emission component.
We note that the fit with the cooling flow model sets tight constraints 
on the existence of a minimum temperature ($\sim$ 2 keV). 
The nominal mass deposition rate in this empirical model is 
$\sim 1900 \mbox{ M}_{\sun} \mbox{ yr}^{-1}$. 
This extremely high $\dot{M}$, in agreement with \textit{Chandra} results 
(Allen et al. 2002), is exceptional for distant cooling core clusters,
and its implications on the evolution of cooling core clusters will
be investigated in a forthcoming paper (Gitti et al. in prep.).


\section{Discussion and Conclusions}

The \textit{XMM-Newton} observation of RX J1347.5$-$1145
confirms that it is, with a luminosity $L_X = 6.0 \pm 0.1 
\times 10^{45} \mbox{ erg s}^{-1}$ (2-10 keV energy band), 
the most X-ray-luminous cluster discovered to date.
RX J1347.5$-$1145 
is a hot cluster (overall temperature: $kT=10.0 \pm 0.3$ keV),
not isothermal: the temperature profile shows the presence of 
a cool core and a decline of the temperature in the outer regions. 
The temperature map reveals a complex structure and identifies a relatively
hot region at radii 50-200 kpc to the SE of the main X-ray peak. 
This hot region is found at the same position as the subclump seen in 
the X-ray image.
The higher temperature and enhanced X-ray surface brightness in the SE 
quadrant indicate that there is probably an ongoing subcluster merger event.
On the other hand, excluding the data of the SE quadrant
the cluster appears relatively relaxed and the presence of a cooling core,
with an exceptional high value of the mass accretion rate 
($\dot{M} \sim 1900 \mbox{ M}_{\sun} \mbox{ yr}^{-1}$) 
indicates that the cluster could have evolved without any disturbances 
for a relatively long interval. 
Therefore, the dynamical state of RX J1347.5$-$1145
appears very complex and will be presented in detail in a forthcoming
paper (Gitti et al. in prep.).


\begin{acknowledgements}
We thank S.Ettori for his advices
in the spectral analysis and for providing the software required to produce the
X-ray colour map in Fig.4, and T. Erben for supplying the optical image.
We also thank the referee J. Kaastra for helpful comments.
M.G. would like to thank E. Belsole, A. Castillo-Morales, S. Majerowicz,
D. Neumann and E. Pointecouteau for suggestions concerning \textit{XMM-Newton} 
data analysis.
This work was supported by the Austrian Science Foundation FWF under 
grant P15868, {\"O}AD Amad{\'e}e Projekt 18/2003 and
{\"O}AD Acciones Integradas Projekt 22/2003. 
\end{acknowledgements}     


\end{document}